\documentclass[prc,superscriptaddress,unsortedaddress,twocolumn]{revtex4}
\usepackage{graphicx,epsf,bm,amsmath,color}
\def\bh{\mbox{\boldmath $h$}}

\def\bp{\mbox{\boldmath $p$}}

\begin{document}
\title{Pauli rearrangement potential for a scattering state with the interaction\\
in chiral effective field theory}
\author{M. Kohno}
\email[]{kohno@rcnp.osaka-u.ac.jp}
\affiliation{Research Center for Nuclear Physics, Osaka University, Ibaraki 567-0047,
Japan}

\begin{abstract}
The Pauli rearrangement potential given by the second-order diagram is evaluated for
a nucleon optical model potential (OMP) with $G$ matrices of the nucleon-nucleon
interaction in chiral effective field theory. The results obtained in nuclear matter are
applied for $^{40}$Ca in a local-density approximation. The repulsive effect is of the
order of 5MeV at the normal density. The density dependence indicates that the
real part of the microscopic OMP becomes shallower in a central region,
but is barely affected in a surface area. This improves the overall resemblance
of the microscopic OMP to the empirical one.

\end{abstract}

\maketitle

\section{Introduction}
An optical model potential (OMP) embodies a basic character of describing the nucleon
elastic scattering on nuclei. The success of the global parametrization of the
phenomenological OMP with certain geometric parameters and a moderately energy-dependent
strength suggests that a mean field picture holds for
scattering states as for bound nucleons. The OMP consists of the real and imaginary
components. The former is regarded as a mean field similar to a single-particle (s.p.)
potential of the ground state, as a consequence of the interaction between the
incoming nucleon and the nucleons in a target nucleus. The imaginary
part takes care of the escaping of an incident flux to inelastic channels.

Microscopic understanding and an explicit evaluation of the OMP, starting from bare
nucleon-nucleon ($NN$) interactions, have been one of the basic problems in nuclear
physics. Jeukenne, Lejeune, and Mahaux \cite{JLM74, JLM76, JLM77} developed
a nuclear matter approach. The properties of the s.p. potential evaluated in nuclear
matter were discussed and a local-density approximation was used for finite
nuclei. Brieva and Rook \cite{BR77a,BR77b} offered a somewhat different method.
A density- and energy-dependent complex effective $NN$ interaction
was prepared based on $G$ matrices in nuclear matter and was applied to construct a
folding potential for finite nuclei, using their localization method of exchange terms.
This study is the prototype of the subsequent $G$-matrix folding model of the
microscopic OMP.

Recent microscopic calculations by several groups \cite{Amo00,Fur08,Toyo18}
with various realistic $NN$ forces are remarkably successful in accounting
for nucleon-nucleus scattering data. The OMP by Amos \textit{et al.} \cite{Amo00},
the Melbourne group, with the Paris \cite{Paris} or Bonn-B \cite{BonnB} potentials
worked well, in spite of the fact that those interactions fail to reproduce
proper nuclear saturation properties. This implies that the description of
nucleon-nucleus scattering is not sensitive to the saturation properties.
Nevertheless, it is appropriate to employ the framework in which the saturation
properties are realized. Furumoto \textit{et al.} \cite{Fur08}
were concerned with the saturation properties by adding the effects of
phenomenological three-nucleon forces (3NFs) to the Nijmegen extended
soft-core $NN$ potential \cite{ESC1,ESC2}. The Kyushu group \cite{Toyo18} employed
the next-to-next-to-next-to-leading order (N$^3$LO) interaction in chiral effective
field theory (ChEFT) \cite{EGM05} with including the effect of the
next-to-next-to-leading order (N$^2$LO) 3NFs \cite{TNF02} in the normal-ordering
prescription \cite{Koh13}. These calculations remain in the lowest-order in the
Brueckner expansion for the OMP. The next-order contribution was discussed
in the early stage of microscopic studies \cite{JLM76}. It was recognized that
the contribution of the second-order rearrangement process becomes smaller
with increasing the incident energy, and therefore the actual incorporation of
this contribution has been left out in the recent microscopic
OMPs \cite{Amo00, Fur08, Toyo18}.
For a deeper understanding of the microscopic OMP, however, it is worthwhile to
revisit the issue of the Pauli rearrangement for the OMP.

The concept of the rearrangement energy was presented by Brueckner and Goldman
\cite{BG60} in the development of the Brueckner theory \cite{BET71}.
Its important role stems from the strong correlations with the $NN$ forces having
short range singularities and Pauli blocking effects. The rearrangement potential
plays a decisive role to reproduce ground state properties of nuclei, which has been
popularized as the potential generated through the derivative of the density-dependent
terms of effective $NN$ interactions in a density-dependent Hartree-Fock description
of nuclei. The density dependence originates partly from the Pauli effects in the
Brueckner theory. Another source of the density dependence is the contribution
of the 3NFs. The main effect of the 3NFs itself can be regarded as the
rearrangement effect due to the Pauli principle acting in the process which includes
the excitation of non-nucleonic degrees of freedom \cite{Koh13}.

In the present article, the contribution of the second-order Pauli rearrangement
diagram for a nucleon scattering state is calculated first in symmetric nuclear matter and
then its implication in finite nuclei is discussed in a local-density approximation.
The treatment is intended to extend the microscopic derivation of the
OMP \cite{Toyo18,Toyo15,Mino16} starting with the $NN$ and 3NF forces parametrized
in ChEFT \cite{EGM05,TNF02}. As is shown in this article,
the inclusion of the Pauli rearrangement potential improves the correspondence
between the microscopic OMP and the empirical one.

Holt \textit{et al.} \cite{Hol13} reported a microscopic calculation of the OMP
in symmetric nuclear matter, at second order in perturbation theory.
Their choice of ChEFT N$^3$LO $NN$ and N$^2$LO 3NF interactions with the
cutoff scale of $\Lambda\simeq 2.5$ fm$^{-1}$ allows the use of the
perturbative framework. The $G$-matrix calculation in the present article
should give quantitatively similar results to theirs.
Here, the results with the larger cutoff scale are also shown.

In Sec. II, basic expressions are given to the Pauli rearrangement
potential in the second order. Numerical results in symmetric nuclear matter
using ChEFT interactions are presented in Sec. III A. The implication of the result
in finite nuclei is discussed in Sec. III B, using a simple local-density
approximation. The conclusions follow in Sec. IV. 

\section{Rearrangement potential}
Ladder correlation is essential for the $NN$ interaction in nuclei to
regularize short-range singularities of the bare $NN$ force. The correlation
naturally depends on the nuclear structure through Pauli effects and the
change of the nucleon propagator. The $G$-matrix equation in the framework
of the Brueckner theory \cite{BET71} properly takes care of these effects, which is
written as:
\begin{equation}
 G(\omega)| ij\rangle =v|ij\rangle +v\frac{Q}{\omega -H_0} G(\omega)|ij\rangle,
\label{eq:geq}
\end{equation}
where $|ij\rangle$ specifies a two-nucleon state, $\omega $ is a sum of
s.p. energies $\omega=e_i +e_j$, and the Hamiltonian $H_0$ is
given by the sum of the kinetic energies and the nucleon s.p. potentials,
$H_0=t_i +U_i+t_j +U_j$. Once the short-range repulsive part is regularized,
the resulting interaction, the $G$ matrix, qualifies for a mean-field description
of low energy nuclear properties. The interaction between an incoming nucleon
and a nucleon in a target nucleus is also considered in the similar framework. 

The microscopic OMP is assigned in the lowest order to the folding potential of the
$G$ matrices with respect to the target wave functions. The method commonly
used \cite{BR77b} is to prepare density- and energy-dependent complex effective
interactions on the basis of nuclear matter calculations, and to apply them in the
folding procedure to finite nuclei.

\begin{figure}[bht]
\centering
 \includegraphics[width=0.4\textwidth,clip]{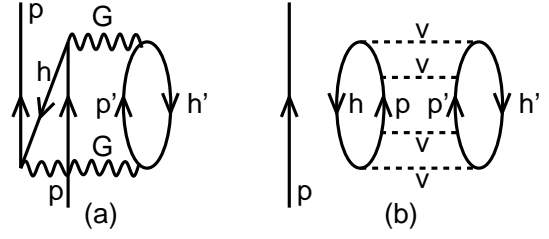}
 \caption{(a) Pauli rearrangement diagram in the second order. (b) Illustration of
the Pauli blocking in a ground-state ladder correlation.}
\label{fig:rea1}
\end{figure}

The next-order contribution, the second-order process in terms of the $G$ matrix,
is shown in Fig. 1(a). This arises as Pauli blocking of the ground state energy of the
target nucleus due to the incoming nucleon, Fig. 1(b). The degree of the importance
of this Pauli blocking effect reflects the importance of the ladder correlation in the
ground state. 
Taking spin and isospin average for the particle state $\bp$, the rearrangement
potential from the diagram Fig. 1(a) is evaluated as 
\begin{align}
 U_{rear.}(\bp)=& -\frac{1}{2} \sum_{\sigma_{p'},\tau_{p'}}
 \sum_{\sigma_{h'},\tau_{h'}} \sum_{\sigma_h,\tau_h} \sum_{\bp'} \sum_{\bh,\bh'}
\notag \\
 & \times \frac{|\langle \bp\bp'|G| \bh\bh' \rangle_A|^2}{e_{\bh}
 +e_{\bh'}-e_{\bp}-e_{\bp'}},
\label{eq:rea}
\end{align}
where $\bh$ and $\bh'$ stand for occupied states (in nuclear matter $|\bh|\le k_F$ and
$|\bh'|\le k_F$ with $k_F$ being the Fermi momentum) and $\bp'$ represents
an unoccupied state (in nuclear matter $|\bp'|>k_F$). The suffix $A$ of the
matrix element denotes antisymmetrization. The summation over $\bp'$
is redundant because of the momentum conservation $\bp+\bp'=\bh+\bh'$.
The potential $U_{rear.}(\bp)$ is apparently real and positive. 
Partial wave expansion is introduced in evaluating $G$ matrices in nuclear matter,
and an angle average is commonly used for the Pauli operator $Q$ and the propagator
in Eq. (\ref{eq:geq}). The angle average is also used in the present calculation of Eq. (\ref{eq:rea}).
If the $G$ matrices are parameterized as a local interaction in a functional form,
such as Gaussian, the partial wave expansion is not necessary. The comparison of
the calculations with and without the angle average for such a parameterized
effective interaction shows that the angle average works very well.

\begin{figure}[thb]
\centering
 \includegraphics[width=0.35\textwidth,clip]{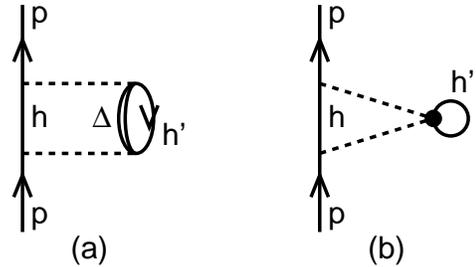}
 \caption{Example of the Pauli blocking process of the 2$\pi$ exchange
3BF: (a) Fujita-Miyazawa type and  (b) ChEFT contact term.}
\label{fig:3nf}
\end{figure}

As the Pauli effect, the rearrangement potential is similar
to the contribution of 3NFs, e.g. the process including $\Delta$-isobar excitation typically.
The excitation of nonnucleonic degrees of freedom that is usually implicit in the
$NN$ potential should be Pauli-blocked in the nuclear medium, Fig. \ref{fig:3nf}(a).
Repulsive effects of this suppression plays an important role to quantitatively
reproduce nuclear saturation properties \cite{Koh13}. The contributions of the
3NFs for describing nucleon-nucleus and nucleus-nucleus scattering problems
have been discussed in recent years \cite{Toyo18}, using the N$^2$LO 3NFs
in ChEFT. The process is illustrated in Fig. \ref{fig:3nf}(b). However,
the estimation of the analogous contributions of the diagram of Fig. \ref{fig:rea1}(a) has
not been considered. The comparison between the contributions of Fig. \ref{fig:rea1}(a)
and Fig. \ref{fig:3nf}(b) is discussed in Sec. III B.

\section{Numerical results}
\subsection{Results in nuclear matter}
The $NN$ interaction in ChEFT parametrized by Epelbaum \textit{et al.} \cite{EGM05}
is used as the bare force. The effects of 3NFs \cite{TNF02} are included in
a normal-ordering prescription. The $G$-matrix calculations in nuclear matter
in the lowest-order Bruckner theory are reported in Ref. \cite{Koh13}, in which
the parameters of the contact 3NFs, $c_D$ and $c_E$, are adjusted so as to
reproduce reasonably well nuclear matter saturation properties. The application
of these $G$ matrices to describing nucleon-nucleus and nucleus-nucleus scattering
processes is presented in Ref. \cite{Toyo18,Mino16}.

\begin{figure}[bth]
\centering
 \includegraphics[width=0.4\textwidth,clip]{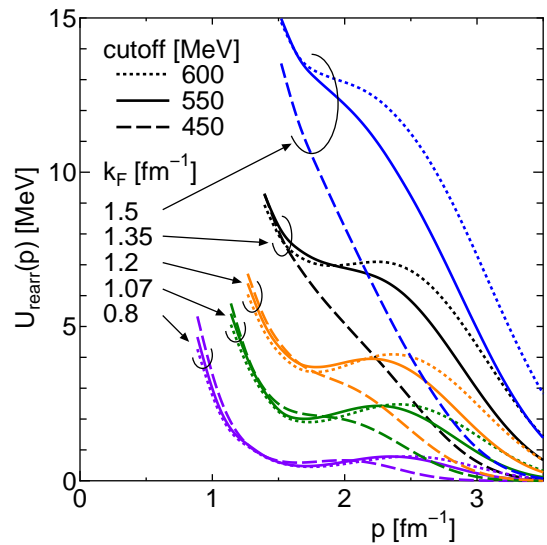}
 \caption{Momentum dependence of the Pauli rearrangement potential in symmetric
nuclear matter with various Fermi momenta from 0.8 to 1.5 fm$^{-1}$. ChEFT
interactions \cite{EGM05} including 3NF effects are employed with three different
cutoff scales; 450, 550, and 600 MeV, respectively.}
\label{fig:reap-p}
\end{figure}

Figure \ref{fig:reap-p} shows calculated results of the Pauli rearrangement potential
in symmetric nuclear matter as a function of the nucleon momentum $p$ for five
cases of the Fermi momentum; 0.8, 1.07, 1.2, 1.35, and 1.5 fm$^{-1}$ which correspond
to the nucleon density as $0.21\rho_0$, $0.5\rho_0$, $0.70\rho_0$, $\rho_0$,
and $1.37\rho_0$, respectively, with the normal density
being $\rho_0=\frac{2}{3\pi^2}(1.35)^3=0.166$ fm$^{-3}$.

\begin{figure}[bht]
\centering
 \includegraphics[width=0.4\textwidth,clip]{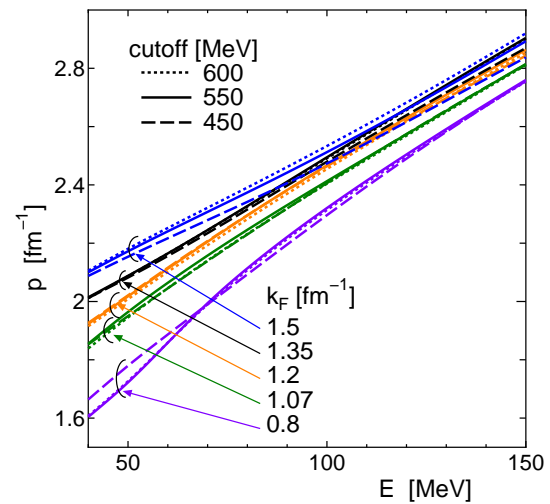}
 \caption{Relation between the energy $E$ and the momentum $p$ in symmetric
nuclear matter specified by Eq. (\ref{eq:pe}).}
\label{fig:ptoene}
\end{figure}

The Pauli rearrangement potential decreases fast as the momentum $p$ increases.
Instead of the momentum $p$, the s.p. energy $E$ of the nucleon is
more adequate to specify the nucleon state in the nuclear medium to describe
nucleon-nucleus scattering, which is related to the momentum $p$ by the
relation including a s.p. potential $U_{LO}(p,k_F)$:
\begin{equation}
 E= \frac{\hbar^2}{2m} p^2 +U_{LO}(p,k_F),
\label{eq:pe}
\end{equation}
where the suffix $LO$ means the lowest-order s.p. potential in nuclear matter.
As explained in Ref. \cite{Koh13}, the following prescription is adopted for $U_{LO}(p)$
when the effective two-body interactions $V_{12(3)}$ deduced from 3NFs by
a normal-ordering prescription are incorporated.
\begin{align}
 U_{LO}(p,k_F)=& \sum_{|\bp'|\le.k_F} \langle \bp\bp'| G \notag \\
 & +\frac{1}{6}V_{12(3)}\left( 1+\frac{Q}{\omega -H}\right) G|\bp\bp'\rangle_A.
\end{align}

\begin{figure}[thb]
\centering
 \includegraphics[width=0.4\textwidth,clip]{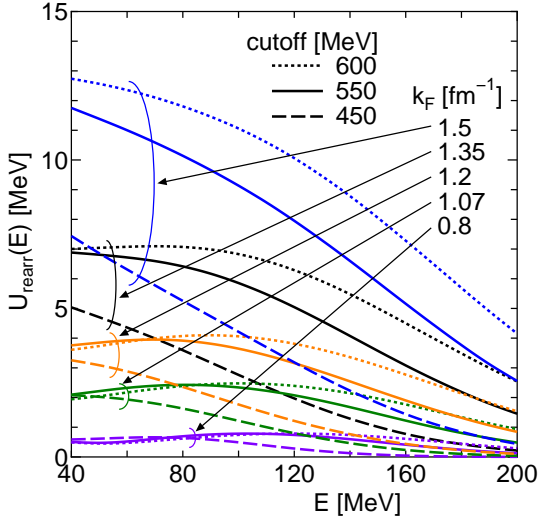}
 \caption{Energy dependence of the Pauli rearrangement potential in symmetric
nuclear matter with various Fermi momenta from 0.8 to 1.5 fm$^{-1}$.  ChEFT
interactions \cite{EGM05} including 3NF effects are employed with three
different cutoff scales; 450, 550, and 600 MeV, respectively.}
\label{fig:reap-E}
\end{figure}

The relation between $p$ and $E$ in nuclear matter with the present
ChEFT interaction is shown in Fig. \ref{fig:ptoene}.
The Pauli rearrangement potential as a function of the energy $E$ is presented
in Fig. \ref{fig:reap-E}. The energy dependence is gentle.
Though some cutoff scale dependence is seen, the strength is not negligible
as far as the density is larger than half the normal density.

\begin{figure}[thb]
\centering
 \includegraphics[width=0.4\textwidth,clip]{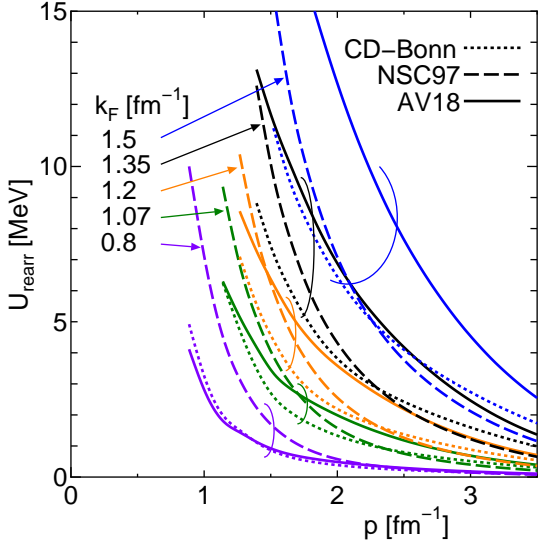}
 \caption{Momentum dependence of the Pauli rearrangement potential in
symmetric nuclear matter calculated with AV18 \cite{AV18}, NSC97 \cite{NSC}
and CD-Bonn \cite{CDB} $NN$ potentials.}
\label{fig:reap-obep}
\end{figure}

It is instructive to carry out similar calculations with other modern $NN$ interactions;
AV18 \cite{AV18}, NSC97 \cite{NSC}, and CD-Bonn \cite{CDB} potentials. Results
are shown in Fig. \ref{fig:reap-obep}. The Av18 potential provides the largest
rearrangement potential among them, the strength of which is comparable to that
of the ChEFT interactions with the cutoff scale of $\Lambda=$ 550 and 600 MeV. 

\subsection{Optical model potential in $^{40}$Ca}
It is worthwhile to consider the rearrangement potential in finite nuclei.
Because the rearrangement potential obtained in nuclear matter is a one-body
quantity, it does not apply in a standard procedure of constructing a
folding potential of effective two-body interactions. In order to estimate the
contribution of the rearrangement potential in a finite nucleus from the result
of the nuclear matter calculation, a simple local-density approximation is employed.

In symmetric nuclear matter, the nucleon density $\rho$ is related to the Fermi
momentum $k_F$ as $\rho=\frac{2}{3\pi^2}k_F^3$. A naive prescription to obtain
a potential in a finite nucleus suggested by the potential $U(p, k_F)$ calculated
in nuclear matter is to replace the Fermi momentum $k_F$ by a local quantity
of $(3\pi^2 \rho(r)/2)^{1/3}$. The momentum $p$ is connected with the incident
energy $E$ by solving Eq. (\ref{eq:pe}).
 
\begin{figure}[thb]
\centering
 \includegraphics[width=0.4\textwidth,clip]{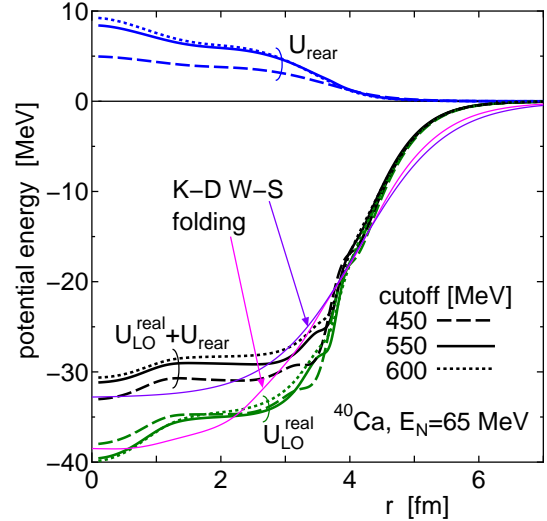}
 \caption{Radial dependence of the real part of the optical model potential
for the nucleon incident energy $E_N=65$ MeV evaluated by a simple
local-density approximation for $^{40}$Ca. The lowest-order potential $U_{LO}^{real}$,
Pauli rearrangement potential $U_{rear}$, and the sum of them are plotted
for three cases of the cutoff scale of the ChEFT interactions. The thin solid
curve labeled ``folding'' represents a localized folding potential \cite{Toyo-p}
obtained with the density-dependent effective two-body interaction parametrized
on the basis of $G$ matrices in nuclear matter \cite{Toyo18}. The thin solid
curve labeled ``K-D W-S'' is a Woods-Saxon potential
by Koning and Delaroche \cite{KD03}.
}
\label{fig:E65}
\end{figure}

\begin{figure}[bht]
\centering
 \includegraphics[width=0.4\textwidth,clip]{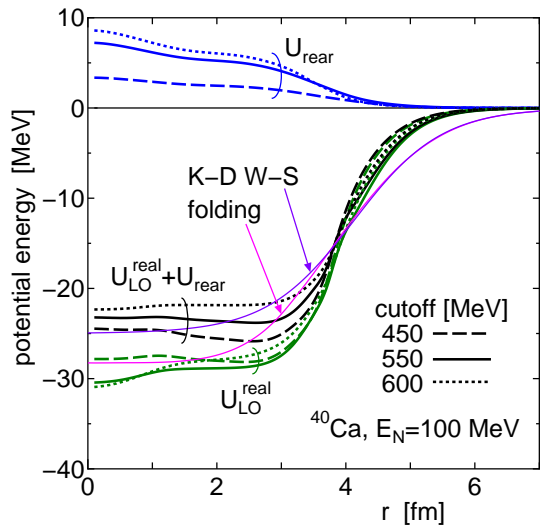}
 \caption{Same as Fig. \ref{fig:E65}, but for $E_N=100$ MeV.}
\label{fig:E100}
\end{figure}

The results for $^{40}$Ca obtained in this prescription are shown in
Figs. \ref{fig:E65} and \ref{fig:E100} for the incident energy of 65 MeV and
100 MeV, respectively. The density distribution of $^{40}$Ca is provided by
Hartree-Fock wave functions with the Gogny D1S effective force \cite{GogD1S}.
The curves marked by $U_{LO}^{real}$ in these figures represent the
lowest-order potential before incorporating the Pauli rearrangement contribution.
It is noteworthy that the shape and the strength of this $U_{LO}^{real}$
corresponds well to that of the folding potential \cite{Toyo-p} calculated
with the density-dependent effective two-body interaction parametrized
on the basis of $G$ matrices of ChEFT interactions \cite{Toyo18} which is
shown by a thin solid curve labeled ``folding'', although the surface thickness is
smaller because of the lack of finite-range effects in the local-density approximation.
This good correspondence assures the usefulness of the local-density
approximation for estimating the contribution of the Pauli rearrangement
effect in finite nuclei.

The Pauli rearrangement effect makes the OMP shallower in a central region
by 5$-$10 MeV, which leads to better correspondence to the depth of
the phenomenological OMP. For comparison, the standard OMP potential
in a Woods-Saxon form by Koning and Delaroche \cite{KD03} is included
in Figs. \ref{fig:E65} and \ref{fig:E100} by a thin solid curve labeled ``K-D W-S''.
In a low-density surface area, the rearrangement effect is small.

\begin{figure}[thb]
\centering
 \includegraphics[width=0.4\textwidth,clip]{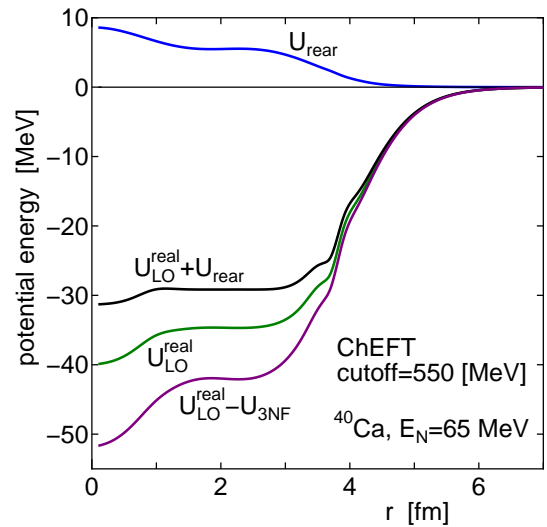}
 \caption{The OMP in which the 3NF contribution is separated is shown for $^{40}$Ca
at $E_N=65$ MeV with the cutoff scale of $\Lambda=550$ MeV.}
\label{fig:ca40full}
\end{figure}

The lowest-order s.p. potential $U_{LO}^{real}(p)$ contains the contribution
of the 3NFs. The chief source of this repulsive contribution is understood
\cite{Koh13} as through Pauli blocking for an excitation process of
nucleon-excited states implicitly present in $NN$ correlations,
as is depicted in Fig. \ref{fig:3nf}. It is interesting to compare the quantitative
contribution of the 3NF with that of the Pauli rearrangement effect.
Figure \ref{fig:ca40full} shows the potential $U_{LO}^{real}(p)-U_{3NF}$ in which
the contribution of the 3NFs is subtracted from $U_{LO}^{real}(p)$.
The 3NF effect makes the potential shallower by about 10 MeV. The Pauli
rearrangement effect provides an additional repulsive potential in a comparable
magnitude. Then the resulting strength of the microscopic OMP in the inner region
becomes closer to the empirical value. Two thin solid curves both in Figs. \ref{fig:E65}
and \ref{fig:E100} indicate that if the effect of the Pauli rearrangement
is incorporated, the microscopic folding potential essentially agrees with the
phenomenological OMP in the whole region.

Differential cross sections and analyzing powers of proton elastic scattering
on $^{40}$Ca, $^{58}$Ni, and $^{208}$Pb at 65 MeV are evaluated in Ref. \cite{Toyo15},
using the microscopic OMP by the density-dependent effective interaction
parametrized on the basis of nuclear matter G-matrix calculation with ChEFT $NN$
and 3$N$ interactions. The results in Fig. 3 of Ref. \cite{Toyo15}
show that the microscopic OMP explains well experimental data and the 3NF effects
are small in spite of the sizable repulsive contribution in a central region as is
presented in Fig. \ref{fig:ca40full}. This is because the nucleon elastic
scattering on nuclei is governed almost by the potential in a low-density
surface area. The same situation applies to the Pauli rearrangement effect.
The successful description of the microscopic OMP
in the literature \cite{Amo00,Fur08,Toyo18} without the rearrangement effect
is scarcely altered by the inclusion of  it.

\section{Conclusions}
A second-order microscopic OMP, a Pauli rearrangement potential, for a
nucleon scattering state is calculated, using ChEFT $NN$ and 3$N$
interactions \cite{EGM05,TNF02}. The second-order potential for a scattering state
is real and positive. The strength of the rearrangement potential in nuclear matter
is 5$-$10 MeV at the normal density. Because the rearrangement potential is obtained
as a one-body potential, it is not straightforward to include this effect
in the standard folding procedure of constructing an OMP for finite nuclei from
effective two-body interactions. In this article, a simple local-density approximation is used to
see qualitatively the possible contribution of the rearrangement potential in finite nuclei.
It is also demonstrated that the repulsive contribution of the Pauli rearrangement
effect is similar in size with the 3NF repulsive contribution.

The density dependence of the rearrangement potential indicates that the OMP
becomes shallower in a central region, but is barely affected in a surface area of low
density. Because the nucleon elastic scattering on nuclei is mainly determined by
the peripheral region, the elastic cross section changes only slightly by the
inclusion of the Pauli rearrangement effect. This explains the success of the recent
microscopic OMP on the basis of realistic $NN$ potentials \cite{Amo00,Fur08,Toyo18},
in spite of taking just the leading order $G$-matrix folding. However, it is
meaningful to observe that the repulsive contribution improves the overall
resemblance of the microscopic OMP to the empirical one.

Nucleus-nucleus scattering has also been described by the $G$-matrix folding
prescription. In this case, higher-density regions participate in the scattering process.
In previous microscopic studies of the nucleus-nucleus elastic scattering \cite{Mino16},
the scattering cross section is still overestimated at larger angles in spite of the
improvement by repulsive and absorptive effects of 3NFs. Then the additional
repulsive rearrangement effect should help to further improve the description,
though the actual implementation of the result in the one-body potential
to the nucleus-nucleus case may not be straightforward.

{\it Acknowledgements.}
This work is supported by JSPS KAKENHI Grant No. JP16K17698. The author
thanks M. Toyokawa for providing him numerical results of the folding potential
in Figs. 7 and 8.

\end{document}